\documentclass{webofc}
\usepackage[mathscr]{euscript}
\usepackage[varg]{txfonts}   
\usepackage{hyperref}
\usepackage{url}
\hypersetup{colorlinks=true,citecolor=blue,urlcolor=blue,linkcolor=blue}
\begin{document}
\title{Nuclear transitions on demand}
%
%

\author{\firstname{Chieh-Jen} \lastname{Yang}\inst{1}\fnsep\thanks{\email{chieh.jen@eli-np.ro}} }

\institute{ELI-NP, ``Horia Hulubei" National Institute for Physics and Nuclear Engineering, 30 Reactorului Street, RO-077125, Bucharest-Magurele, Romania
          }

\abstract{I show a way to tune photo-nuclear cross section effectively and therefore achieve nuclear transitions ``on demand". The method is based on combinatorial enhancement of multiphoton processes under intense conditions. Taking advantage of recent advances in high-power laser systems (HPLS) and nuclear structure calculations, efficient control of nuclear transitions up to E4 in multipolarity can be reached today. The same idea can be extended to the search for rare transitions and hidden states, which applies to the $\gamma$-beams generated from conventional sources as well.
}
\maketitle
\section{The key question }
\label{intro}

One main obstruction which prevents femto-technology (manipulation of nuclear excited states) can be illustrated in Fig~\ref{fig1}. Formulating in simple terms, one can ask the following:
\begin{figure}[h]
\centering
\includegraphics[width=10cm,clip]{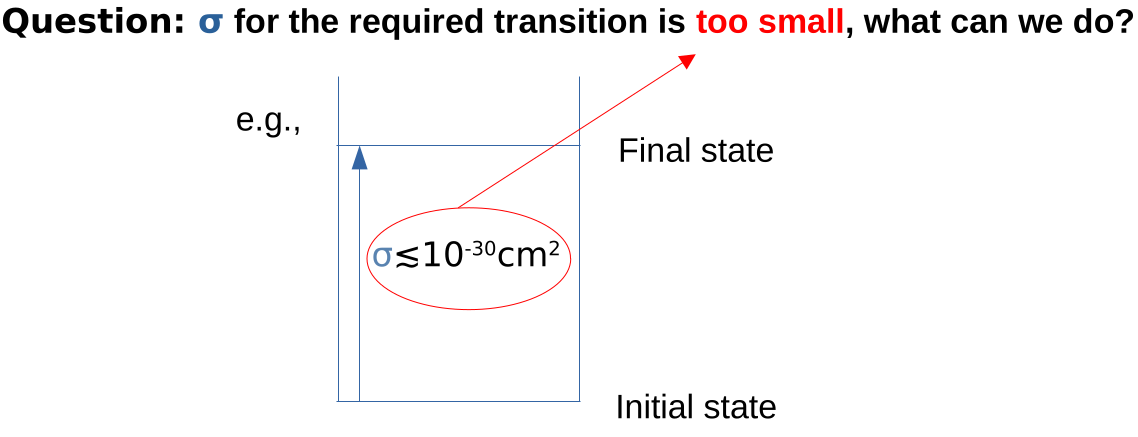}
\caption{Illustration of a photo-nuclear transition.}
\label{fig1}       
\end{figure}

Dimensionally, the cross section $\sigma$ scales as $\sim \pi r^2$, where $r$ is the size of the system. Thus, going down from the nano- to the femto- scale, one expects the cross section to be suppressed by $(10^{6})^2$. Indeed, the typical electromagnetic transitions in atoms/molecules are of the order of Mb, whereas the same is much smaller in the nuclear case. Even in the most pronounced case, e.g., giant dipole resonance, the cross section is of the order of mb. This $10^9-10^{12}$ suppression factor compared to photonics at nanoscales has greatly hindered the practical applications of nuclear photonics.
Solving this obstacle would greatly improve the utilization of nuclear energy, bringing a tremendous breakthrough in human technology.

\section{Main idea}
\label{sec-1}
Defining $\sigma_{if}$ the cross section of a transition from state $|i\rangle$ to $|f\rangle$, the yield ${Y_{f}}$ corresponding to Fig.~\ref{fig1} is
\begin{equation}
    Y^{\text{1 to 1}}_f=\iint [I_{\Gamma_f} \sigma_{if} N z] dt d\mathscr{A},
\label{eq1}    
\end{equation}
where $I_{\Gamma_{f}}$ is the number of photons per area ($\mathscr{A}$) per unit time ($t$) within the energy interval that corresponds to the width of state $|f\rangle$. $N$ and $z$ are the number density of the state $|i\rangle$ in the target nuclei and its thickness, respectively.
Note that Eq.~(\ref{eq1}) describes an \textit{one-to-one} process, where each state $|i\rangle$ of the target interacts with an incoming particle (photon in this example) to reach $|f\rangle$. In this case, the yield $Y_f$ is linearly proportional to $I_{\Gamma_{f}}$.
Thus, if one insists on the process specified in Fig.~\ref{fig1} and Eq.~(\ref{eq1}), then it is impossible to increase the yield except for feeding the target with more photons.

\begin{figure}[h]
\centering
\includegraphics[width=12cm,clip]{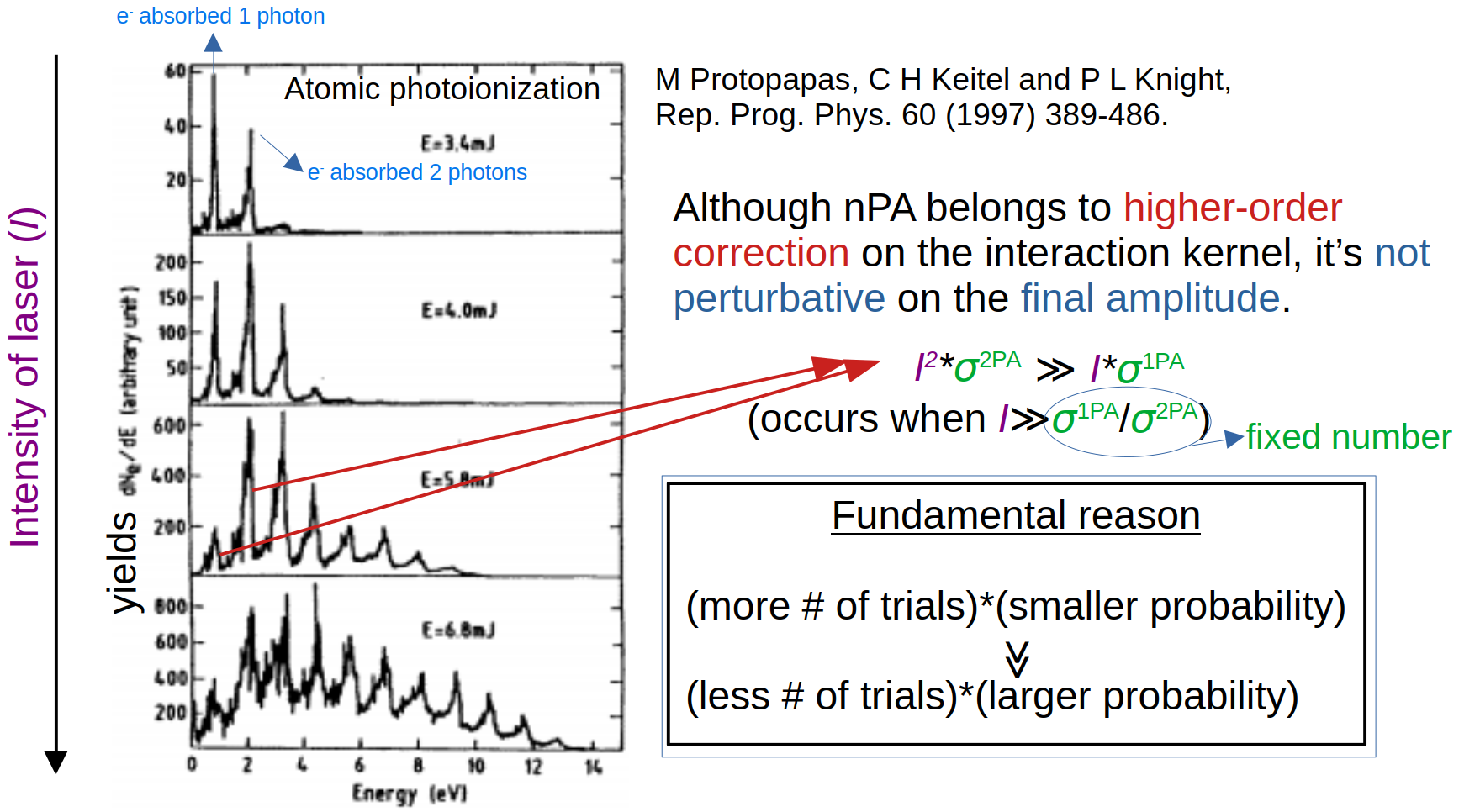}
\caption{The nPA yield for different intensity of incoming photons. Here data of atomic photonionization is presented, with photon intensity increases from top to bottom panel. The peak energy flux $\mathcal{P}\equiv IE_{\omega}\approx10^{13}$ W/cm$^2$ (where $E_{\omega}\approx1.2$ eV for Nd:YAG laser). The plot is extracted from Fig.~6 of Ref.~\cite{Protopapas1997}. }
\label{figa}       
\end{figure}

\begin{figure}[h]
\centering
\includegraphics[width=10cm,clip]{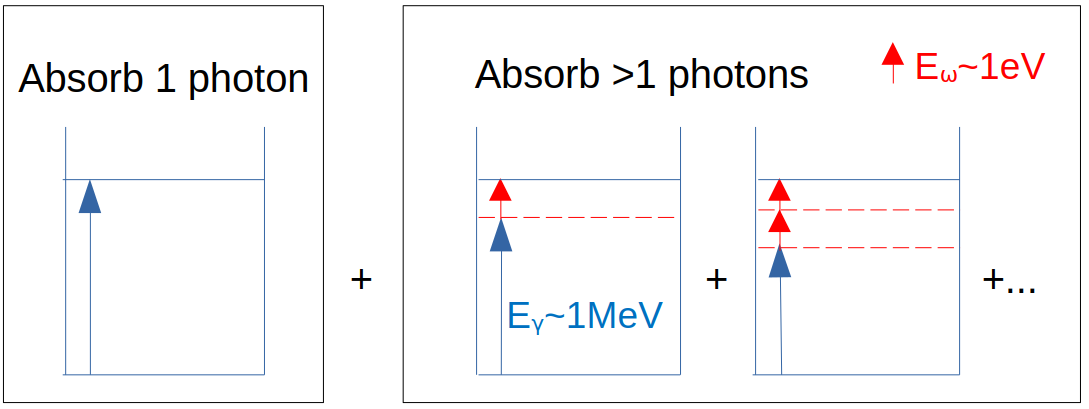}
\caption{Illustration of nPA processes (gaps between states are not to scale). Here nPA start at an initial state and goes through a series of virtual sates (denoted as red-dashed lines) to reach the final state. }
\label{fig2}       
\end{figure}

However, Eq.~(\ref{eq1}) is not the complete story of the transition from state $|i\rangle$ to $|f\rangle$. Whenever symmetry does not prohibit, \textit{more-to-one} processes as described in Fig.~\ref{fig2} can occur. Thus, the yield actually consists of 
\begin{equation}
    Y_f=\sum_n y_f^{nPA}=y^{1PA}_f+y^{2PA}_f+...,
\label{eq2}    
\end{equation}
where $y_f^{1PA}$ corresponds to Eq.~(\ref{fig1}), and 
\begin{equation}
y^{2PA}_f=\iint [I_{\Gamma_f}(I_{\Gamma_f}-1) \sigma^{2PA}_{if} N z] dt d\mathscr{A}.
\label{eq3}    
\end{equation}
Here $y_f^{2PA}$ is the yield of two-photon absorption (2PA)~\cite{GM}, which is the next simplest/probable mechanism to achieve $|i\rangle \rightarrow |f\rangle$. The factor $I_{\Gamma_f}(I_{\Gamma_f}-1)$ originates from a combinatorial enhancement~\cite{Yang:2019hkn,Yang:2021vxa,10.1117/12.3056128}, which becomes $I_{\Gamma_f}^2$ when $I_{\Gamma_f}\gg1$. The 2PA cross section, $\sigma^{2PA}_{if}$, is often expressed in unit of [cm$^{4}$s].\footnote{The unit of $\sigma^{nPA}_{if}$ is [length]$^{2n}$[time]$^{n-1}$.} It can be evaluated from second-order perturbation theory on the transition kernel and is a rare event compared to single-photon absorption. But since both $\sigma_{if}$ and $\sigma^{2PA}_{if}$ have a fixed value, as the intensity increases, the contribution of $I_{\Gamma_f}^2\sigma^{2PA}_{ij}$ must exceed $I_{\Gamma_f}\sigma_{ij}$. Note that the promotion of this \textit{two-to-one} event does not violate any self-consistency in the perturbation theory. It happens simply because one fed a rare event with a much larger number of trials, which is achieved by the aforementioned combinatorial enhancement under the intense regime with the same incoming flux (i.e., $I_{\Gamma_f}^2\gg I_{\Gamma_f}$). As shown in Fig.~\ref{figa}, this phenomenon has been observed experimentally in atomic photoionization, with yields that correspond to theoretical predictions~ (see e.g., Fig.~6 of Ref.~\cite{Protopapas1997}, the basic formula Eq. (3.1) and further references therein).      

Moreover, the photons that participate in nPA need not be of the same energy. In inhomogeneous two-photon absorption, Eq.~(\ref{eq3}) becomes
\begin{align}
y^{2PA}_f=&\iint [I_{\Gamma_f}I_2 \sigma^{2PA}_{if} N z] dt d\mathscr{A},\notag \\ 
\equiv&\iint [I_{\Gamma_f}\sigma^{2PA}_{eff} N z] dt d\mathscr{A},
\label{eq4}    
\end{align}
where $I_2$ can belong to low-energy photons (e.g., eV-level photons directly from a high-power laser). $\sigma_{eff}^{2PA}\equiv I_2\sigma_{if}^{2PA}$ can then be seen as a \textit{tunable} effective cross section "felt" by the $\gamma$-photons with intensity $I_{\Gamma_f}$. The beauty of this scheme is that the $\gamma$-photons (represented by the blue arrows in Fig.~\ref{fig2}) are responsible of filling the energy gap of nuclear transitions, whereas the optical photons (represented by the red arrows in Fig.~\ref{fig2}), though negligible in energy in the nuclear scale, boost the combinatorial choices and enhance the effective cross section felt by the $\gamma$-photons. In this way, not only can a rare transition (as described in Fig.~\ref{fig1}) be greatly enhanced, it actually allows one to \textit{adjust} the effective cross section as desired and therefore achieves \textit{nuclear transition on demand}. The concept is directly expansible to n-photon processes (nPP) so that a combination of absorptions and stimulated emissions can be carried out, which then allows one to cover a wider range of the angular momentum quantum number ($J^{\pi}$) difference between $|i\rangle$ and $|f\rangle$.

\section{Current status and technical limitations}
Note that the fantastic idea of enhancing the effective cross section by supplying $\gamma$-photons with optical photons from a laser is not new. The same idea was introduced in Refs.~\cite{PhysRevLett.42.1397,PhysRevC.20.1942} in 1979, where it was estimated that $\sigma^{2PA}_{eff}$ of an originally E2 transition can be boosted up to $10^{-26}$ cm$^2$ (10 mb) when the energy flux of the supplied optical photon $\mathcal{P}_2\equiv E_{\omega}I_2= 10^{10}$ W/cm$^2$ (where $E_{\omega}\sim 1$ eV).
Despite the physical significance of the idea, it is still practically of limited use. First, $\gamma$-photons need to be accurately adjusted to an energy so that $E_{\gamma}+E_{\omega}$ matches the energy gap (with precision up to the width of $|f\rangle$) for 2PA to occur, whereas the energy gap varies between states across nuclei. Second, since the optical photons mostly need to resonate with the initial or the final state itself, this means that the virtual state will have the same $J^{\pi}$ as the state to which it resonates (see, e.g., Ref.~\cite{PhysRevC.20.1942} for details). Thus, the enhancing mechanism through 2PA can only be realized in a final state separated from the initial state by at most E2 in multipolarity, which greatly limits practical applications\footnote{A multipolarity gap greater than E2, though can still be enhanced, will have an effective cross section $\lesssim10^{-27}$ cm$^2$ even supplied with a flux of 10 PW optical photons. }.   

The introduction of modern HPLS~\cite{Yoon:19,Yoon:21,Li:18,Yu:18,ur2015eli,Tanaka2020,Cernaianu2025} is a game changer for the aforementioned scheme. By compressing pulses within an extremely tiny space and time ($\sim5$ $\mu$m, $\sim20$ fs), modern HPLS can now reach $\mathcal{P}_2$ up to 10$^{23}$ W/cm$^2$, and therefore provide a much more intense source of optical photons. Moreover, when being irradiated onto an overdense target, laser-matter interaction is capable of generating various intense beams, including the MeV-level $\gamma$ flash. The laser-driven $\gamma$ flash presents an intensive and \textit{continuous} energy spectrum, with up to $\mathcal{P}_2\approx10^{12}$ W/cm$^2$ per eV interval for $E_{\gamma}$ anywhere between $1$ keV $\sim1$ MeV~\cite{PhysRevApplied.13.054024,Gu2018,Xue2020,PhysRevE.104.045206,Heppe2022}. As a result, nPA can be realized with n up to 4 by overlapping the $\gamma$ flash with optical photons supplied by another laser. Further studies show that with state-of-the-art HPLS, it is feasible to enable nuclear transitions that cover a multipolarity gap up to E4 with an effective cross section $\sigma_{eff}\approx10^{-25}$ cm$^2$~\cite{our_iso}. 

\section{A demonstration of 2PA v.s. stepwise pumping}
\begin{figure}[h]
\centering
\includegraphics[width=6cm,clip]{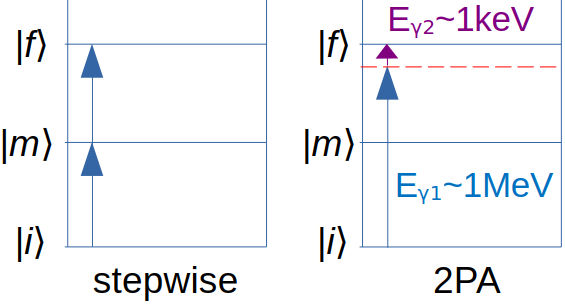}
\caption{Illustration of stepwise pumping processes versus 2PA. Here the stepwise pumping start at state $|i\rangle$ and goes through an intermediate physical state $|m\rangle$ to the final state $|f\rangle$, while 2PA goes through the virtual sates (denoted as red-dashed lines) instead of $|m\rangle$ to reach the final state. Gaps between states are not to scale.}
\label{fig3}       
\end{figure}

 To further demonstrate the superiority of nPA over conventional stepwise pumping in the intensity limit, an example is provided as follows.
Consider a nuclear system as depicted in Fig.~\ref{fig3}. Suppose that there exists an intermediate state $|m\rangle$ between the initial and final states $|i\rangle$ and $|f\rangle$, and that one wishes to pump from $|i\rangle$ to $|f\rangle$. 
For stepwise pumping (1PA+1PA), the nucleus then first needs to absorb a photon to reach $|m\rangle$, then absorbs another photon to reach $|f\rangle$ before $|m\rangle$ decays. The yield for the first excitation in the 1PA+1PA process is:
\begin{equation}
    Y_1=\iint [I_{\Gamma_1} \sigma_1 N z] dt dA,
    \label{1st}
\end{equation}
where $I_{\Gamma_1}$ is the number of photons per area ($A$) per unit time ($t$) within the energy interval that corresponds to the width of the intermediate state. $\sigma_1$ is the cross section towards the intermediate state $|m\rangle$. $N$ and $z$ are the number density of the target nuclei and its thickness. As the typical duration of a $\gamma$ flash is 15-50 fs---which is comparable or shorter than the half-life of a typical intermediate nuclear state ($\approx 0.01\sim1$ ps)---one can integrate the above equation (i.e., assuming no intermediate state will decay for the entire duration of the $\gamma$ flash) and obtain
\begin{equation}
    Y_1=X_{\Gamma_1}\sigma_1Nz,
    \label{1sti}
\end{equation}
where $X_{\Gamma_1}$ is the total number of photons from the $\gamma$ flash within energies $E\approx E_{intermediate}\pm \Gamma_{1}/2$, with $\Gamma_1$ the width of the intermediate state.
For the next excitation, one applies Eq.~(\ref{1st}) again with $N$ replaced by the number density of the first excited state, $\frac{Y_1}{V}$ ($V$ is the volume), and the index in subscript 1 replaced by 2. The final yield of 1PA+1PA is:
\begin{equation}
    Y_{1PA+1PA}=\iint [ I_{\Gamma_2} \sigma_2 \frac{Y_1}{V} z] dt dA,
    \label{2st}
\end{equation}
Assuming that all photons are parallel with each other and enter orthogonally into the target with an impact area $A_{in}$, then $V=A_{in}z$.  
On the other hand, the yield from 2PA (by the 1 MeV + 1 keV combination)\footnote{Here I consider all photons are provided solely by a single $\gamma$ flash generated by laser-matter interaction without additional optical photon supplied and assume the lower end of the spectrum $E_{\gamma}\approx1$ keV.} is:
\begin{equation}
    Y_{2PA}^{1MeV+1keV}=\iint [ I_{\Gamma_2} \sigma_{eff}^{2pa} N z] dt dA.
    \label{2pa}
\end{equation}
Thus, the ratio between 1PA+1PA and 2PA is
\begin{equation}
    \frac{Y_{1PA+1PA}}{Y_{2PA}^{1MeV+1keV}}=\frac{X_{\Gamma_1}\sigma_1\sigma_2}{A_{in}\sigma_{eff}^{2pa}}.
    \label{2pa}
\end{equation}
For simplicity, let us assume zero angular divergence but take a conservative value of the photon intensity from the $\gamma$ flash, i.e., $10^{6}$ $\gamma$-photons per eV interval are generated within an area of $10\times10$ $\mu$m$^2(\equiv A_{in})$ per shot (duration $\sim50$ fs). Assume a typical nuclear intermediate state, that is, $t_{1/2}\approx1$ ps, which corresponds to a width $\approx10^{-4}$ eV. This factor weakens the number of available $\gamma$-photons from $X_{\Gamma_{eV}}=10^6$ (per eV) to $X_{\Gamma_1}=10^2$ (per $10^{-4}$ eV) for 1PA+1PA to occur. Substituting $X_{\Gamma_1}= 10^{2}$ and assuming $\sigma_1=\sigma_2=10^{-24}$ cm$^2$, $\sigma_{eff}^{2pa}=10^{-37}$ cm$^2$,\footnote{This value can be obtained by substituting the aforementioned conditions into Eq.~(9) of Ref.~\cite{our_iso} for $A\approx200$ nuclei.} and $A_{in}=10^{-6}$ cm$^{2}$, one obtains the ratio of Eq.~(\ref{2pa})$\approx 10^{-3}$ (note that without the $10^4$ suppression from $\Gamma_1$, yields from 1PA+1PA would be comparable to 2PA (1MeV+1keV) under the given $\gamma$-flash). Then, considering all possible combinations of photon energies within 2PA leads to another $\approx10^2$ enhancement (although there are $10^6/2$ combinations, but those with their virtual state further away from the final state are suppressed as they are "out of resonance"). Thus, with an intense HPLS-driven $\gamma$ flash, the yield of 2PA could already exceed stepwise pumping by a $\approx 10^5$ factor. Note that this is achieved \textit{without} supplying the $\gamma$ photons with optical photons from another laser, which will further enhance the nPA yield greatly.    

\section{Conclusion and new opportunities in probing rare events}
\begin{figure}[h]
\centering
\includegraphics[width=12cm,clip]{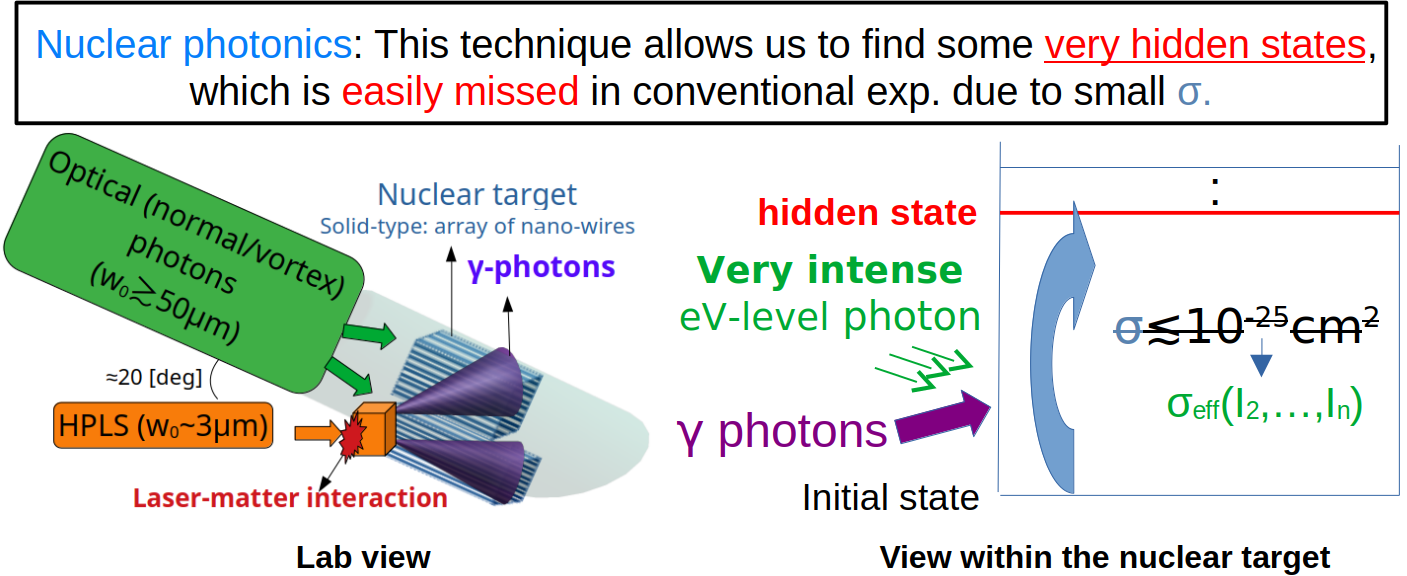}
\caption{Illustration of new opportunities opened via nPA. }
\label{fig4}       
\end{figure}

The proposed scheme not only enables nuclear transition on demand, but opens new opportunities in probing rare events.
For example, the tunable effective cross section may allow us to discover some very hidden state in the nuclei as well, as illustrated in Fig.~\ref{fig4}. In this case, one supplies very intense eV-level photons to boost the effective cross section "felt" by the $\gamma$ photons (which can be from any source), so that they can reach an excited state originally very hidden due to being associated with a very small single-photon cross section. Moreover, it leads to a new idea for realizing nuclear gamma-ray lasers (graser)~\cite{graserour}.

In summary, the ability to manipulate nuclear states will bring tremendous practical applications, including
medical imaging and therapy, energy storage and generation, the treatment of radioactive materials,
and precision measurement of fundamental physics both within and beyond the standard model.
Collaboration and innovation between nuclear and laser-plasma communities is required to unlock the immense possibilities that lie ahead.

\section*{Acknowledgments}
I would like to express my deep gratitude to Bruce Barrett (1939-2025), who not only hired me at my first postdoc, but also introduced to me the fascinating nuclear many-body world.
I thank Paolo Tomassini, Klaus Spohr, Vojtech Horny, Domenico Doira, Septimiu Balascuta and Bogdan Corobean for stimulating ideas and providing complemental knowledge from the laser-plasma side.  
This work was supported by the Extreme Light Infrastructure Nuclear Physics (ELI-NP) Phase II, a project co-financed by the Romanian Government and the European Union through the European Regional Development Fund - the Competitiveness Operational Programme (1/07.07.2016, COP, ID 1334);  the Romanian Ministry of Research and Innovation: PN23210105 (Phase 2, the Program Nucleu), the ELI-RO grant Proiectul ELI-RO/RDI\_2024\_AMAP, ELI-RO\_RDI\_2024\_LaLuThe, ELI-RO\_RDI\_2024\_SPARC and ELI10/01.10.2020 of the Romanian Government; the European Union, the Romanian Government and the Health Program, within the project ``Medical applications of high-power lasers - Dr. LASER"; SMIS Code: 326475 and the IOSIN funds for research infrastructures of national interest.
We acknowledge EuroHPC Joint Undertaking for awarding us access to Karolina at IT4Innovations (VŠB-TU), Czechia, under project number OPEN-34-63 and EHPC-REG-2023R02-006 (DD-23-157), and CINECA HPC access through PRACE-ICEI standard call 2022 (P.I. Paolo Tomassini).

%

%
%
\bibliography{iso} 

\end{document}